\documentclass[12pt, preprint]{aastex}

\shorttitle{Eddington Accretion and QSO Emission Lines at $z\sim$2}
\shortauthors{Yuan \& Wills}

\begin{document}

\title{Eddington Accretion and QSO Emission Lines at $z\sim$2}

\author{Michael Juntao Yuan and Beverley J. Wills}
\affil{Astronomy Department, University of Texas, Austin, TX 78712}

\begin{abstract}

Broad Absorption Line (BAL) QSOs have been suggested to be youthful super-accretors based on their powerful radiatively driven absorbing outflows and often reddened continua. To test this hypothesis, we observed near IR spectra of the H$\beta$ region for 11 bright BAL QSOs at redshift $z\sim$2. We measured these and literature spectra for 6 BAL QSOs, 13 radio-loud and 7 radio-quiet non-BAL QSOs. Using the luminosity and H$\beta$ broad line width to derive black hole mass and accretion rate, we find that both BAL and non-BAL QSOs at $z\sim$2 tend to have higher $L/L_{\rm Edd}$ than those at low $z$ -- probably a result of selecting the brightest QSOs. However, we find that the high $z$ QSOs, in particular the BAL QSOs, have extremely strong \ion{Fe}{2} and very weak [\ion{O}{3}], extending the inverse relationship found for low $z$ QSOs. This suggests that, even while radiating near $L_{\rm Edd}$, the BAL QSOs have a more plentiful fuel supply than non-BAL QSOs. Comparison with low $z$ QSOs shows for the first time that the inverse \ion{Fe}{2} -- [\ion{O}{3}] relationship is indeed related to $L/L_{\rm Edd}$, rather than black hole mass.  

\end{abstract}
\keywords{quasars: emission lines --- quasars: general --- galaxies: active --- galaxies: evolution --- accretion, accretion disks --- black hole physics}

\section{INTRODUCTION}

Broad Absorption Line (BAL) QSOs show very broad blueshifted absorption troughs indicating outflows up to $\sim0.1c$, with covering fractions $\ga 0.2$ \citep{hew03}, and soft X-ray hydrogen column densities often exceeding $10^{23} cm^{-2}$ \citep{gallagher02}. These outflows carry a significant fraction of the accretion power and there is evidence that they are radiatively driven \citep{arav97}, suggesting that BAL QSO black holes accrete close to the Eddington limit.  Many BAL QSOs, especially those with low ionization absorption troughs, are reddened \citep{low88, becker00, hall02}, suggesting that BAL QSOs may be young QSOs ejecting a birth-cocoon of dusty gas to reveal the unobscured optical QSO \citep{fabian99,haehnelt98}. With the discovery of many BAL QSOs at $z\sim2$, we are able to explore this idea by comparing Eddington accretion ratio ($L/L_{\rm Edd}$) directly with fueling indicators in BAL and non-BAL QSOs.

We can estimate $M_{\rm BH}$ and hence $L/L_{\rm Edd}$ for $z\sim2$ QSOs using observed continuum luminosity and FWHM~H$\beta$ in the following way: If QSO Broad Line Region (BLR) gas motion is dominated by gravity, the black hole mass ($M_{\rm BH}$) can be calculated from the typical orbital radius of BLR clouds and their typical velocity: $M_{\rm BH} = v^2 r / G$. BLR radius $r$ can be represented by BLR size ($R_{\rm BLR}$) derived from H$\beta$ reverberation mapping: $R_{\rm BLR}$ can be estimated from continuum luminosity at rest wavelength 5100\AA: $R_{\rm BLR} \propto L_\lambda(5100$\AA$)^{0.66}$ \citep{peterson99, kaspi00, vestergaard02}; $v$ can be represented by H$\beta$ full width at half maximum (FWHM). The $M_{\rm BH}$ so derived agree with the $M_{\rm BH}$ derived independently from the $M_{\rm BH}$ - bulge luminosity and $M_{\rm BH} \propto \sigma^4$ relationships for nearby QSOs, where $\sigma$ is the velocity dispersion of the bulge \citep{laor98, gebhardt00, mclure02}.

The fueling could be indicated by correlated emission line properties. One of the strongest sets of relationships among QSO properties describes the increasing strength of optical \ion{Fe}{2} with decreasing [\ion{O}{3}]~$\lambda$5007, increasing steepness of the soft X-ray continuum, and decreasing width and stronger blue wing of the broad H$\beta$ emission line \citep[][ hereinafter BG92]{grupe99, shang03, bg92}.  Here we call this set of relationships Boroson \& Green Eigenvector 1 (BGEV1). It has been suggested that BGEV1 is driven by $L/L_{\rm Edd}$ for the following reasons.

\begin{itemize}

\item For a given luminosity, narrower H$\beta$ corresponds to smaller $M_{\rm BH}$ and higher $L/L_{\rm Edd}$.

\item By analogy with Galactic black hole binary systems, it has been suggested that steeper X-ray spectra indicate higher $L/L_{\rm Edd}$ \citep{pounds95,done95}. Near-Eddington accretion results in geometrically thick accretion disks which produce excess soft X-ray photons \citep[e.g.][]{tanaka96}. 

\item Strong optical \ion{Fe}{2} emission and weak [\ion{O}{3}] emission indicate high optical depth, high density nuclear gas that may provide an abundant fuel supply for near-Eddington accretion. Optically thick gas reduces ionizing photons that reach the [\ion{O}{3}] narrow line region \citep[e.g.][]{bg92}. Strong \ion{Fe}{2} suggests that this gas may have been metal-enriched by star formation. The few known BAL QSOs at low $z$ tend to have narrow H$\beta$, weak [\ion{O}{3}] and strong \ion{Fe}{2} emission \citep{turnshek97, weymann91} suggesting high $L/L_{\rm Edd}$ and an abundant fuel supply.

\end{itemize}

Most BALs are discovered in QSOs with $z > 1.5$, where the broad \ion{C}{4} troughs are shifted into the optical window. For those redshifts the H$\beta$ region is shifted to the near infrared. We therefore performed near infrared spectroscopy, taking advantage of new QSO surveys that do not use UV selection and yield many more BAL QSOs than previously recognized \citep{becker00, reichard03}.  In this letter, we investigate whether BAL QSOs have extreme BGEV1 properties, using FWHM~H$\beta$ and continuum luminosity to calculate $L/L_{\rm Edd}$, and then compare $L/L_{\rm Edd}$ with \ion{Fe}{2} and [\ion{O}{3}] strengths. Detailed discussion of the observations, reductions and analysis, with results from a larger data set, will be published later.

\section{SAMPLES AND DATA}

\subsection{Samples and Observations}

Our high redshift ($z\sim2$) sample consists of 37 QSOs. We observed 11 BAL QSOs from the Large Bright QSO Survey, the FIRST Bright QSO Survey and \citet{weymann91}. QSOs of similar redshift and luminosity were observed by \citet[ hereinafter M99a]{m99a}. We included their observations having acceptable signal-to-noise ratio, to make a total sample of 17 BAL QSOs, 13 radio-loud (RL) QSOs and 7 non-BAL radio-quiet (RQ) QSOs. We and M99a chose the brightest optically-selected QSOs at $z\sim2$.  

Our new spectra were obtained with the CGS4 near infrared spectrograph on the United Kingdom Infra-Red Telescope (UKIRT). The detector was the 256x256 InSb NICMOS array. With the 40~l~mm$^{-1}$ low resolution grating, a slit width of 1.2\arcsec\, yielded a resolution of 3 pixels as measured from Ar and Xe comparison lamp spectra ($\sim600$~km~s$^{-1}$ in J band and $\sim750$~km~s$^{-1}$ in H band).

The two-dimensional spectral images were reduced using UKIRT's ORAC data reduction pipeline. We used standard IRAF tasks to optimally extract the spectra and calibrate their wavelength scales. Atmospheric features were removed and spectral shape calibration was done by division by the observed spectra of F-G stars using a temperature implied by the spectral type. The F-G stars and the wavelength comparison lamps were observed within 0.1 airmasses of the QSOs. Our reduced spectra have signal-to-noise ratios of from 10 to 20 per 3-pixel bin near H$\beta$.

\subsection{Spectral Decomposition}

Measuring line and continuum properties in the H$\beta$ region is complicated by the blending of many broad optical \ion{Fe}{2} lines. We used SPECFIT \citep{kriss94} within IRAF to deblend spectral components for both our UKIRT and the M99a spectra. We decomposed each spectrum into a powerlaw continuum, broad and narrow H$\beta$ components, a broad H$\gamma$ component, a narrow-line [\ion{O}{3}] doublet and \ion{Fe}{2} emission blends.

The fixed parameters were the power-law continuum parameters (slope and normalization estimated by eye), the [\ion{O}{3}]~$\lambda\lambda$4959,5007 doublet ratio of 2.94, the H$\gamma$/H$\beta$ ratio of 0.36, and the narrow H$\beta$ to [\ion{O}{3}]~$\lambda$5007 ratio of 0.1 \citep{veilleux87}. Except for some M99a spectra with known broad [\ion{O}{3}] lines, the narrow H$\beta$ and [\ion{O}{3}] lines were represented by single Gaussians with width equal to the instrumental resolution (BG92).  The \ion{Fe}{2} emission blends were represented by the BG92 I~Zw~1 template. All broad lines including \ion{Fe}{2} were assumed to have the same Gaussian profile. Rest wavelength ratios were constrained except for some McIntosh et al. objects with shifted [\ion{O}{3}] lines \citep{m99b}. Free parameters were the intensities of [\ion{O}{3}], broad H$\beta$, the \ion{Fe}{2} blends, and broad line widths.

\begin{figure}
\plotone{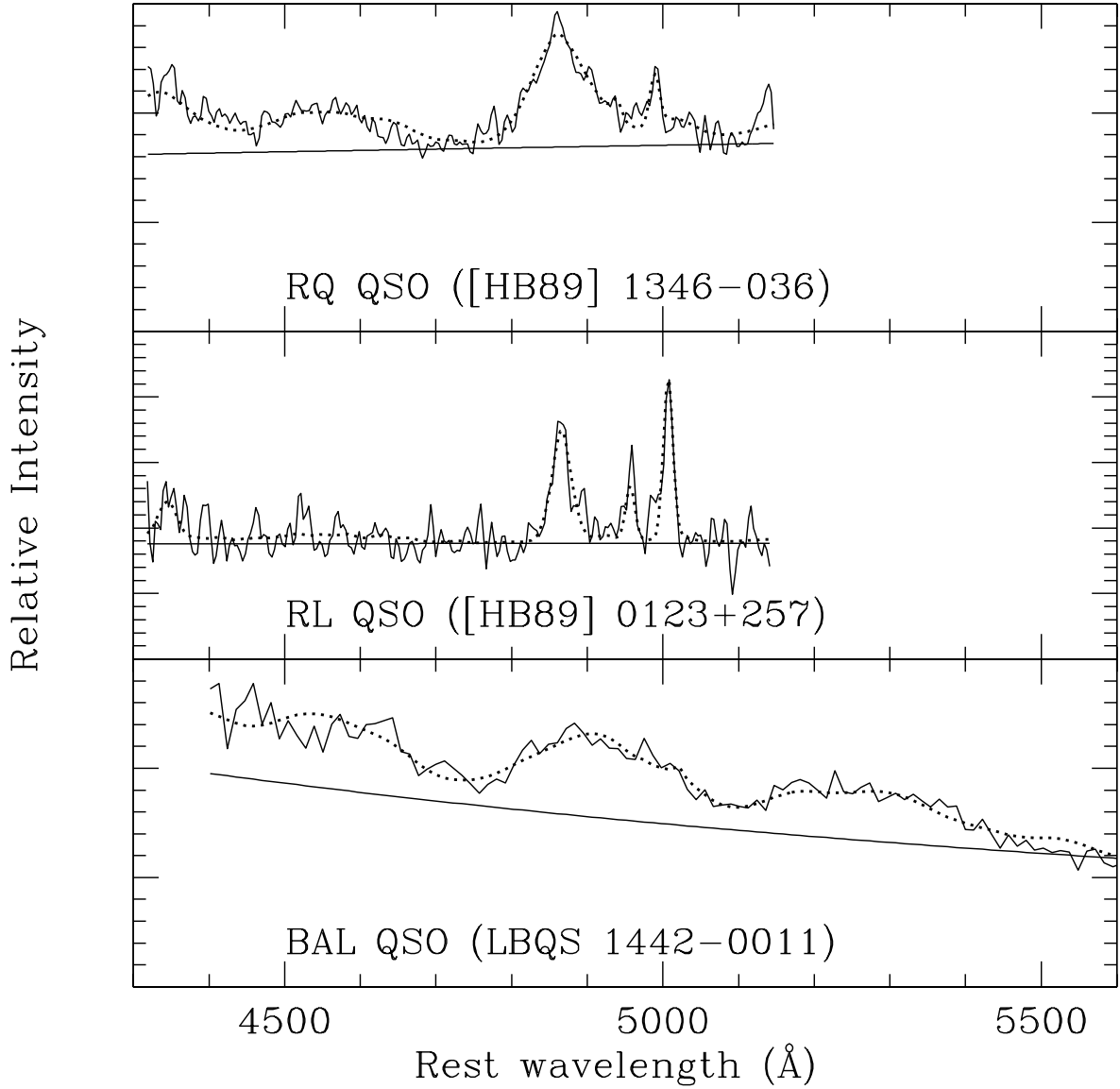}
\caption{Typical observed spectra (solid lines) and fitted models (dotted lines) for high redshift QSOs. The underlying solid line is the fitted continuum.
\label{spectra}}
\end{figure}

In many cases, since a single Gaussian profile was inadequate to represent the broad H$\beta$ line, we adopted the following procedure after running SPECFIT. We first subtracted all fitted components except broad H$\beta$ and smoothed the remaining H$\beta$ by fitting multiple Gaussian profiles, each at least as wide as the instrumental resolution, then measured the integrated flux and FWHM of the smoothed profile. We tested our measurement method on BG92 spectra, finding no systematic difference from their results. For objects without measurable [\ion{O}{3}] or \ion{Fe}{2} lines, we estimated $3\sigma$ upper limits based on the signal-to-noise ratio and spectral resolution of each spectrum. The typical rms uncertainties of our measurements for the high redshift QSOs are 10\% for [\ion{O}{3}]~$\lambda$5007 equivalent width (EW) for objects with detected [\ion{O}{3}], 10\% for FWHM~H$\beta$, and about 20\% for \ion{Fe}{2}/H$\beta$ ratios. Figure~\ref{spectra} shows the fits for three typical spectra. The BAL QSO spectrum is from our new UKIRT data. The RL and RQ QSO spectra are from our refitting of the M99a data.

\clearpage

\clearpage

\subsection{Continuum Luminosities}

To calculate $M_{\rm BH}$ and $L/L_{\rm Edd}$, we need $L_{\lambda}$ at rest frame wavelengths of 5100\AA\, and 3000\AA\, to estimate the BLR size, and the bolometric luminosity: $L_{\rm bol}=8.3\cdot\lambda L_{\lambda}(3000$\AA$)$ \citep{laor00}. BAL QSOs tend to have reddened continua compared with optically selected non-BAL QSOs, so the $L_{\lambda}$ were calculated using H band magnitudes from the the 2MASS survey\footnote{\url{http://www.ipac.caltech.edu/2mass/}}, assuming $f_{\nu} \propto \nu^{-0.5}$. Using the most recent BAL QSO colors derived from the Sloan Digital Sky Survey \citep{tolea02}, we estimate a generous upper limit for extinction at the observed H band correponding to $\sim$0.8 mag for $z\sim2$, corresponding to under-estimates of black hole mass and Eddington ratio by factors of $<10^{-0.2}$ and $<10^{-0.1}$ respectively. Such small corrections would not change our results significantly.  We assumed a cosmological model with $H_0 = 70~km~s^{-1}~Mpc^{-1}$, $\Omega_m=0.3$ and $\Omega_\Lambda=0.7$ \citep{freedman02}. Table~\ref{datatable} lists the line measurements and calculated $L / L_{\rm Edd}$ for our high $z$ QSOs. The complete table is available electronically.

\clearpage

\begin{deluxetable}{cccccc}
\tablecolumns{6}
\tablewidth{0pc}
\tablecaption{$z \sim$ 2 QSO properties \label{datatable}}
\tablehead{
\colhead{Name} & \colhead{Type} & \colhead{EW~[\ion{O}{3}]} & \colhead{\ion{Fe}{2}/H$\beta$} & \colhead{$M_{\rm BH}$} & \colhead{$L / L_{\rm Edd}$} \\
\colhead{} & \colhead{} & \colhead{\AA (rest)} &
\colhead{} & \colhead{$10^9M_{\odot}$} & \colhead{}}
\startdata
0046+0104 & BAL &  11.3 &  3.0 & 3.3 & 0.78 \\
0052+0101 & RQ  &   6.4 &  5.6 &  5.1 & 0.9 \\
0052+0140 & RQ  &   5.7 &  2.9 &  0.7 & 3.2 \\
0126+2559 & RL  &  20.9 &  1.7 &  0.4 & 6.76 \\
0157+7442 & RL  &  19.7 &  1.3 & 19.4 & 0.42 \\
0228$-$0337 & RL &   23.9 &  1.2 &  1.9 & 1.07 \\
0228$-$1011 & BAL &   7.0 &  3.3 &  9.5 & 0.26 \\
0424+0204 & RL  &  50.6 &  1.3 & 15.5 & 0.30 \\
0427$-$1302 & RL  &  29.9 &  1.0 &  8.6 & 0.47 \\
0555+3948 & RL  &  14.9 & $<$1.8 &  1.9 & 1.71 \\
0724+4159 & BAL &  $<$5.6 &  7.4 &  1.6 & 1.91 \\
0841+7053 & RL  &   4.8 &  5.1 &  4.0 & 1.86 \\
0845+3420 & BAL &   5.6 &  4.2 & 16.4 & 0.18 \\
0913+3944 & BAL &  $<$3.0 &  8.7 &  0.1 & 9.48 \\
0934+3153 & BAL &  $<$1.8 &  3.1 &  5.1 & 0.80 \\
1013+0851 & BAL &  $<$1.3 &  4.6 & 14.3 & 0.39 \\
1054+2536 & BAL &  $<$2.7 &  4.2 &  2.5 & 1.07 \\
1106$-$1821 & RQ &   12.9 &  2.3 &  4.4 & 1.0 \\
1225+2235 & RQ  &   8.9 &  1.9 & 35.8 & 0.3 \\
1228+3128 & RL  &  $<$0.9 &  5.1 & 40.8 & 0.39 \\
1231+0725 & RL  &   8.3 & $<$0.7 &  3.5 & 1.04 \\
1233+1304 & BAL &  $<$6.2 &  3.1 &  8.3 & 0.41 \\
1234+1308 & BAL &  $<$3.0 &  2.4 &  2.8 & 1.04 \\
1249$-$0559 & BAL &  $<$2.3 &  5.0 & 22.9 & 0.51 \\
1250+2631 & RQ &   10.6 &  1.5 & 11.4 & 1.2 \\
1311$-$0552 & BAL &  $<$1.3 &  6.0 &  4.8 & 0.94 \\
1333+1649 & RL &   18.7 &  1.7 & 17.5 & 0.58 \\
1348$-$0353 & RQ &    4.3 &  3.3 &  3.9 & 1.3 \\
1418+0852 & RQ &    8.2 &  2.3 &  4.6 & 1.0 \\
1420+2534 & BAL &  $<$4.9 & 11.1 &  3.0 & 1.07 \\
1436+6336 & RL  &  11.6 &  1.1 & 16.4 & 0.31 \\
1445+0129 & BAL &  $<$3.7 &  2.6 &  1.1 & 1.13 \\
1445$-$0023 & BAL &  $<$4.2 &  9.5 &  8.0 & 0.17 \\
1451$-$2329 & RL &   22.4 &  1.4 &  4.8 & 1.20 \\
1516+0029 & BAL &  $<$2.1 &  1.4 &  6.7 & 0.48 \\
2215$-$1744 & BAL &  20.3 &  1.7 & 15.4 & 0.25 \\
2312+3847 & RL &   28.4 &  5.3 &  5.4 & 0.63 \\
\enddata
\end{deluxetable}

\clearpage

\section{RESULTS AND DISCUSSION}

\subsection{Are BAL QSOs Extreme Accretors?}

\begin{figure}
\plotone{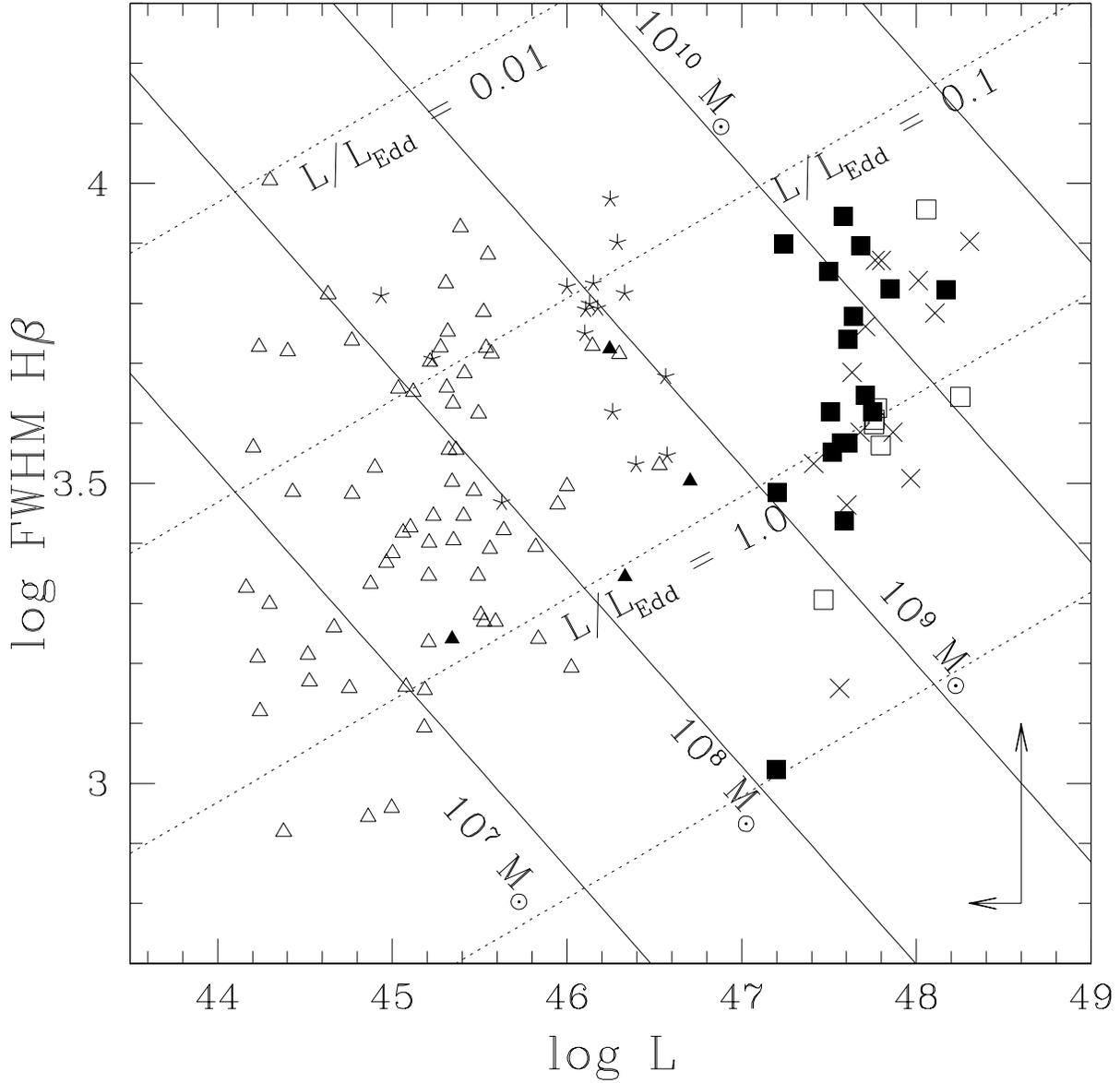}
\caption{FWHM~H$\beta$ versus bolometric luminosity. Lines of constant $M_{\rm BH}$ and constant $L/L_{\rm Edd}$ are shown. See Fig.~\ref{O3Fe} for symbols. The arrows show possible corrections to the high $z$ RL QSO data for velocity projection and continuum beaming.
\label{LbolHbFWHM}}
\end{figure}

Figure~\ref{LbolHbFWHM} plots FWHM~H$\beta$ against bolometric luminosity. Lines of constant $M_{\rm BH}$ and $L/L_{\rm Edd}$ are indicated. High redshift QSOs (large symbols) are compared with the 87 low redshift BG92 QSOs from which the original BGEV1 relationships were derived (small symbols). Different sub-samples of QSOs are distinguished.  The $L/L_{\rm Edd}$ for our flat-spectrum RL QSOs might be overestimated if the continuum is beamed \citep{padovani92}, and FWHM~H$\beta$ may underestimate the true virial speed as a result of viewing disk motions pole-on \citep[e.g.][]{krolik01}. The brightest QSOs at $z\sim 2$ are the most luminous, introducing a bias toward the highest $L/L_{\rm Edd}$ and $M_{\rm BH}$. The high $z$ BAL QSOs, while having high $L/L_{\rm Edd}$, are not clearly different from the high $z$ non-BAL QSOs, even after applying the orientation corrections.

\clearpage

\clearpage

\subsection{The \ion{Fe}{2} versus [\ion{O}{3}] Anti-correlation}

\begin{figure}
\plotone{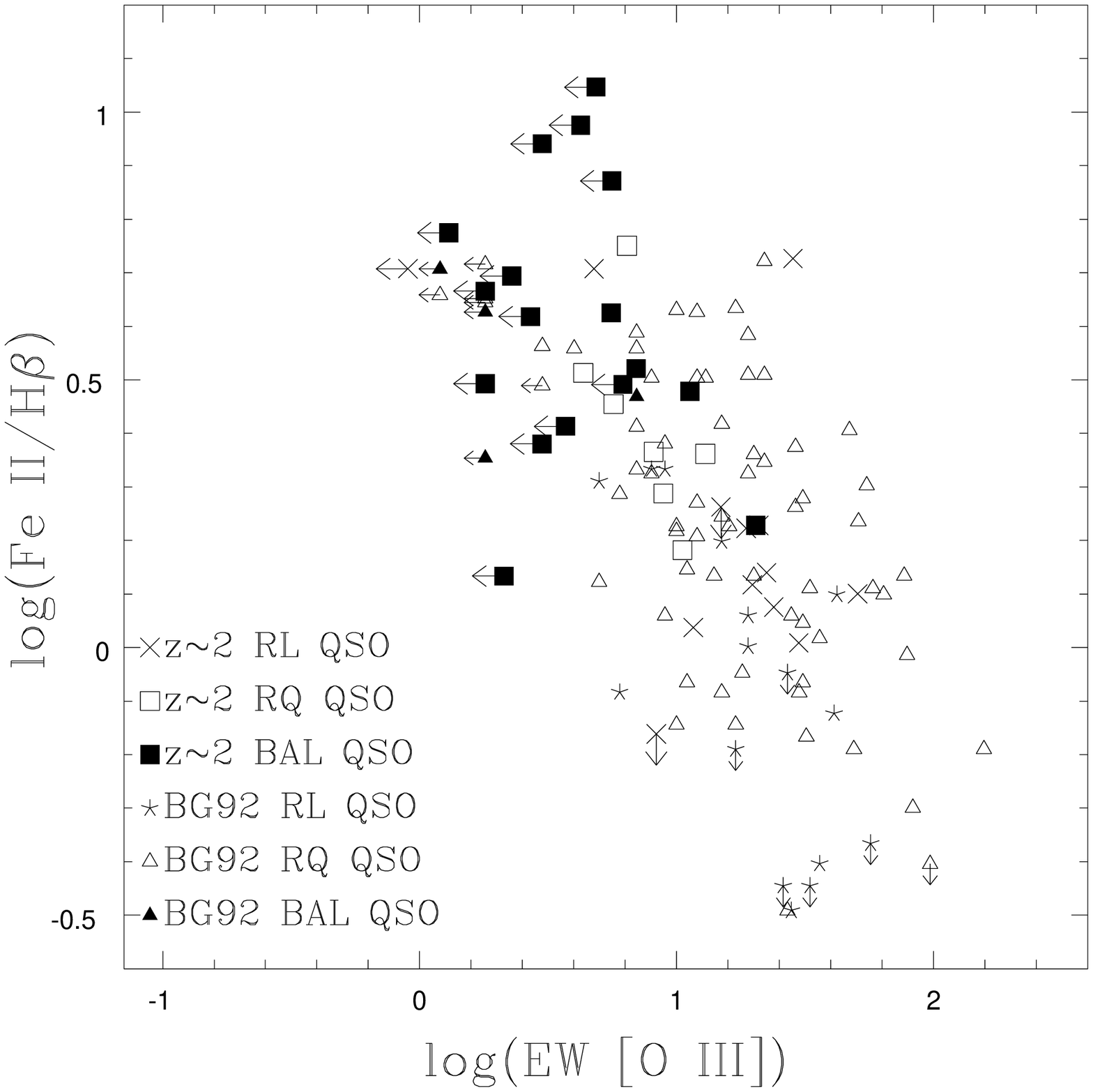}
\caption{Intensity ratio \ion{Fe}{2}/H$\beta$ versus EW~[\ion{O}{3}]~$\lambda$5007.
\label{O3Fe}}
\end{figure}

Figure~\ref{O3Fe} shows that high $z$ QSOs, in particular the BAL QSOs, extend the inverse relationship between \ion{Fe}{2}/H$\beta$ and EW~[\ion{O}{3}] to stronger \ion{Fe}{2} and weaker [\ion{O}{3}]. Thirteen out of seventeen high $z$ BAL QSOs do not even have detectable [\ion{O}{3}] emission lines. Taking into account the [\ion{O}{3}] upper limits, the generalized Kendall's tau correlation coefficient indicates that the 1-tailed probability for the \ion{Fe}{2} versus [\ion{O}{3}] correlation to arise by chance is $< 0.01\%$ for either the entire sample or the high $z$ QSOs alone. The trend for low $z$ RL and BAL QSOs to lie at opposite extremes has been noted before (BG92). The same appears to be true among the high $z$ QSOs. The 1-tailed probability of high $z$ BAL QSOs and low $z$ RQ QSOs having the same line strength distribution is $< 0.1\%$ for \ion{Fe}{2}/H$\beta$ and $< 0.1\%$ for EW~[\ion{O}{3}].

\clearpage

\clearpage

\subsection{Eddington Ratio Indicator}
\begin{figure}
\plotone{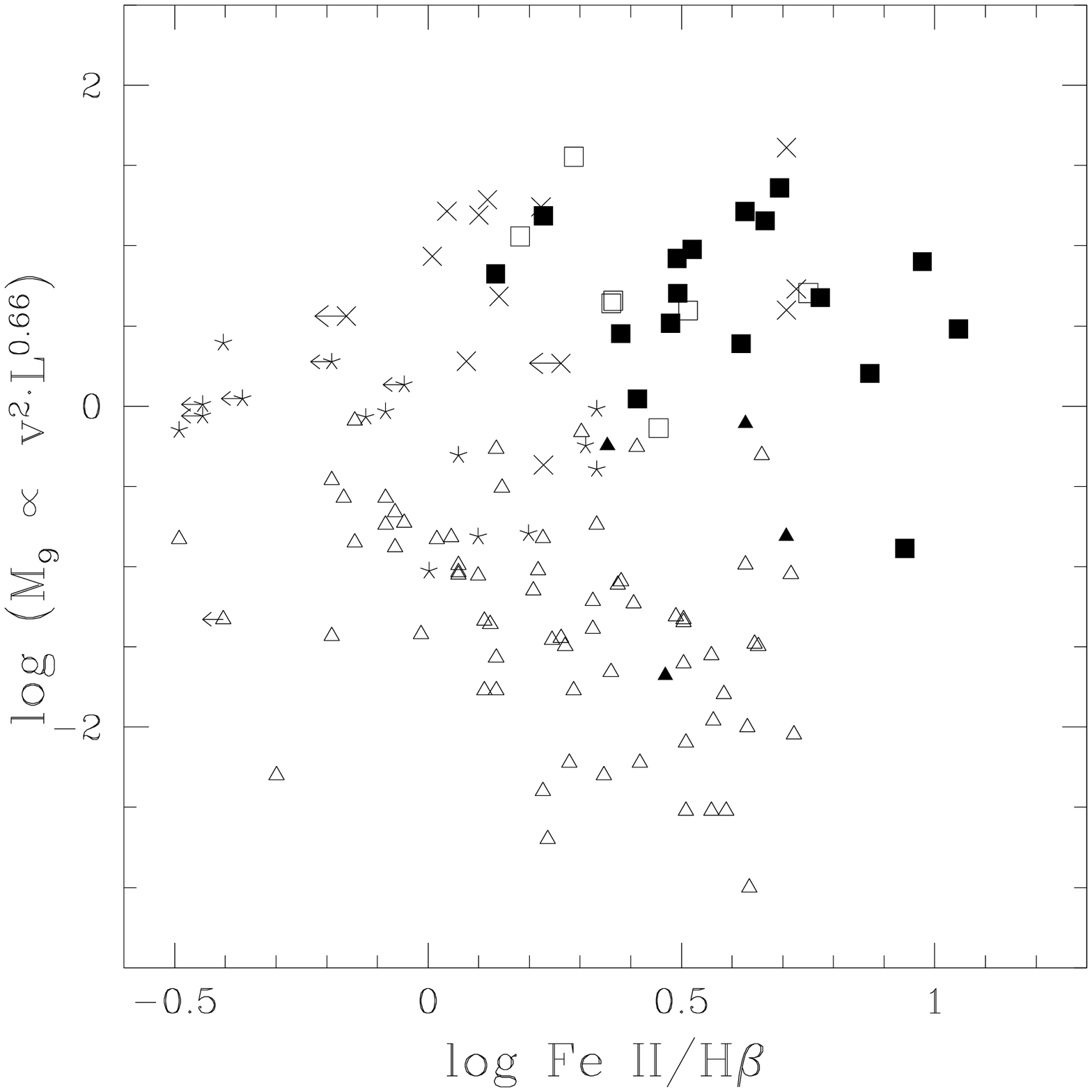}
\caption{$M_{\rm BH}$ versus \ion{Fe}{2}/H$\beta$. See Fig.~\ref{O3Fe} for symbols.
\label{FeM}}
\end{figure}

\begin{figure}
\plotone{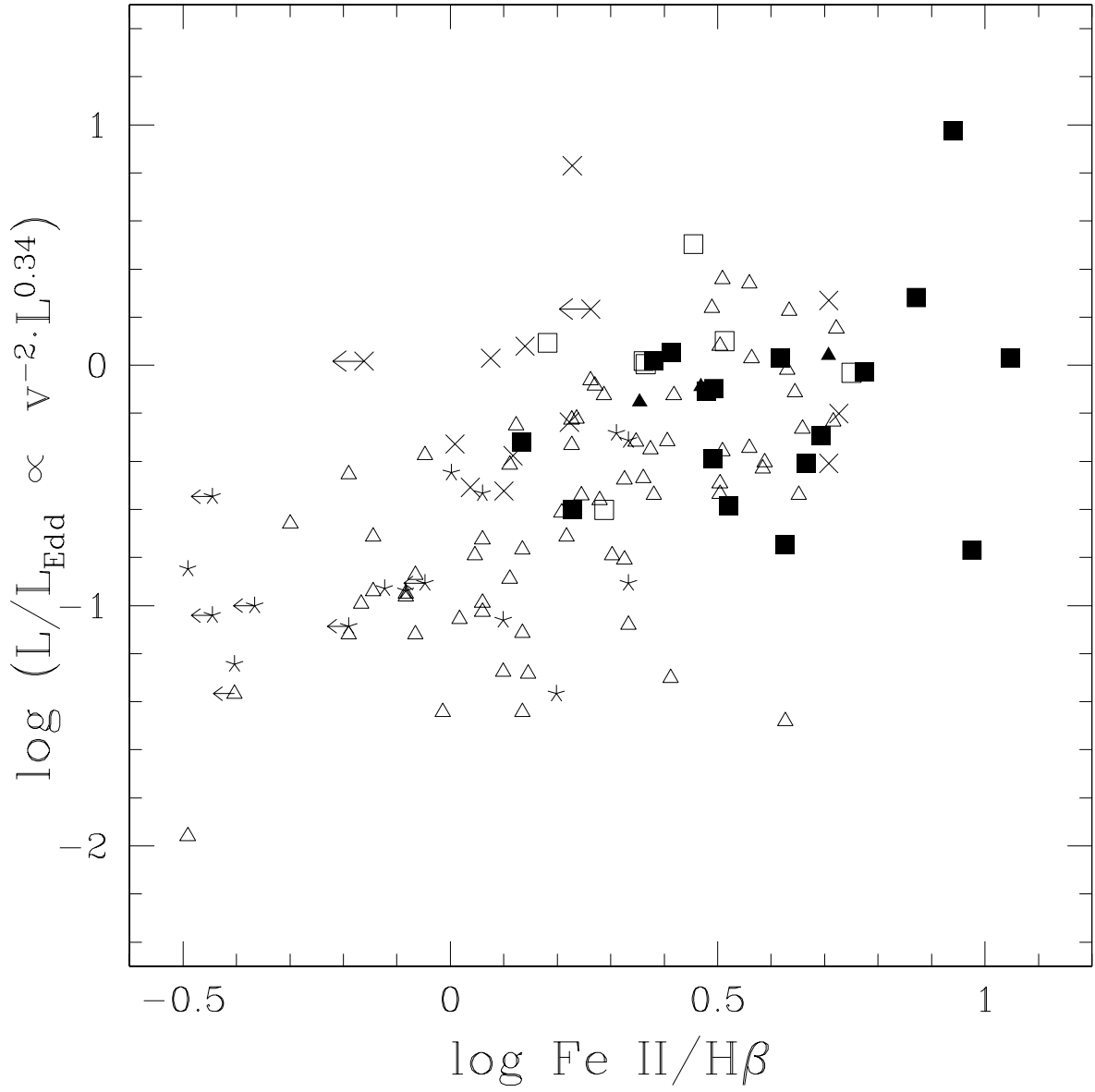}
\caption{$L/L_{\rm Edd}$ versus \ion{Fe}{2}/H$\beta$. See Fig.~\ref{O3Fe} for symbols.
\label{FeMdot}}
\end{figure}

.
Figures \ref{FeM} and \ref{FeMdot} show $M_{\rm BH}$ and $L/L_{\rm Edd}$ versus \ion{Fe}{2}/H$\beta$. In Fig.~\ref{FeM}, the $z\sim2$ QSOs lie at significantly larger $M_{\rm BH}$; in Fig.~\ref{FeMdot}, these high redshift QSOs extend the low $z$ relationship to higher $L/L_{\rm Edd}$. For both high and low $z$ QSOs, the 2-tailed probability of the $L/L_{\rm Edd}$ versus \ion{Fe}{2}/H$\beta$ correlation arising by chance is $< 0.01\%$. The $M_{\rm BH}$ versus \ion{Fe}{2}/H$\beta$ correlation for low $z$ QSOs may simply reflect the inverse dependence of $M_{\rm BH}$ on $L/L_{\rm Edd}$, given a limited luminosity range. There is also an inverse correlation between $L/L_{\rm Edd}$ and EW~[\ion{O}{3}] for both high and low $z$ QSOs (not shown), with most BAL QSOs lying at the weak [\ion{O}{3}] extreme (see Fig.~\ref{O3Fe}). The 2-tailed probability of this correlation arising from unrelated variables is $< 0.01\%$. The above correlations demonstrate that BGEV1 is indeed related to $L/L_{\rm Edd}$ rather than  $M_{\rm BH}$. {\it If BGEV1 properties are $L/L_{\rm Edd}$ indicators, why do high $z$ BAL and non-BAL QSOs have different BGEV1 properties despite their similar $L/L_{\rm Edd}$?} One possibility is that all our high $z$ QSOs belong to the same parent population but BAL QSOs are viewed from special angles. This would imply that BGEV1 emission line properties are affected by orientation. More likely, the BGEV1 emission line relationships may be only indirectly related to $L/L_{\rm Edd}$. They may depend more directly on the availability of fuel around the black hole. Stronger \ion{Fe}{2} and weaker [\ion{O}{3}] correspond to an abundance of cold gas, which could fuel high Eddington ratio accretion. For low $L/L_{\rm Edd}$ objects, which are found mostly in the low $z$ sample, increasing the fuel supply increases the accretion rate. At higher luminosities (high $z$) BAL QSOs may be the youngest QSOs with the most abundant fuel supplies, as indicated by very strong \ion{Fe}{2} and very weak [\ion{O}{3}] emission, but there is not a corresponding increase in accretion ratio because they are unable to radiate at $L > L_{\rm Edd}$ 

\clearpage

\clearpage

\acknowledgements We thank P. Hirst for observing support, D. Wills for comments on a draft, D. H. McIntosh for generously providing H band spectra for high redshift QSOs, T. A. Boroson for spectra of the BG92 sample, and Roc Cutri for providing unpublished 2MASS magnitudes.  The United Kingdom Infra-Red Telescope is operated by the Joint Astronomy Centre on behalf of the U.K. Particle Physics and Astronomy Research Council.

\end{document}